\documentclass[twocolumn,superscriptaddress,nobibnotes,nofootinbib,floatfix,aps,prl,citeautoscript,reprint]{revtex4-1}

\usepackage{graphicx}
\usepackage{epsfig}
\usepackage{subfigure}
\usepackage{color}
\usepackage{latexsym}
\usepackage{amsmath,bm}
\usepackage{dcolumn}           
\usepackage{natbib}
\newcommand{\etal}{\textit{et al.~}}
\newcommand{\ub}{Department of Physics, Universit\"{a}t Basel,
Klingelbergstr. 82, 4056 Basel, Switzerland}
\newcommand{\lsim}{Laboratoire de Simulation Atomistique (L\_Sim), SP2M, INAC, CEA-UJF, 38054 Grenoble Cedex 9, France}
\newcommand{\lpmcn}{Universit\'e de Lyon, F-69000 Lyon, France and 
LPMCN, CNRS, UMR 5586, Universit\'e Lyon 1, F-69622 Villeurbanne, France}
\newcommand{\lsi}{Laboratoire des Solides Irradi\'es and ETSF, \'Ecole Polytechnique, 
CNRS, CEA-DSM, 91128 Palaiseau, France}

\begin{document}

\title{The crystal structure of cold compressed graphite}

\author{Maximilian Amsler}
\affiliation{These authors contributed equally to this work.} \affiliation{\ub}

\author{Jos\'e A. Flores-Livas} \affiliation{These authors contributed equally to this work.} \affiliation{\lpmcn}

\author{Lauri Lehtovaara}
\affiliation{\lpmcn}
\author{Felix Balima}
\affiliation{\lpmcn}
\author{S. Alireza Ghasemi}
\affiliation{\ub}
\author{Denis Machon}
\affiliation{\lpmcn}
\author{St\'ephane Pailh\`es}
\affiliation{\lpmcn}
\author{Alexander Willand}
\affiliation{\ub}
\author{Damien Caliste}
\affiliation{\lsim}
\author{Silvana Botti}
\affiliation{\lsi}
\affiliation{\lpmcn}
\author{Alfonso San Miguel}
\affiliation{\lpmcn}
\author{Stefan Goedecker}
\email{stefan.goedecker@unibas.ch}
\affiliation{\ub}
\author{Miguel A.L. Marques}
\email{miguel.marques@univ-lyon1.fr}
\affiliation{\lpmcn}

\date{\today}

\begin{abstract}
  Through a systematic structural search we found an allotrope of
  carbon with $Cmmm$ symmetry which we predict to be more stable than
  graphite for pressures above 10\,GPa. This material, which we refer
  to as Z-carbon, is formed by pure $sp^3$ bonds and is the only
  carbon allotrope which provides an excellent match to unexplained
  features in experimental X-ray diffraction and Raman spectra of
  graphite under pressure. The transition from graphite to Z-carbon
  can occur through simple sliding and buckling of graphene sheets.
  Our calculations predict that Z-carbon is a transparent wide band
  gap semiconductor with a hardness comparable to diamond.
\end{abstract}

\maketitle

Thanks to the flexibility to form $sp$, $sp^2$ and $sp^3$ bonds,
carbon is one of the most versatile chemical elements.  At ambient
pressure, it is usually found as graphite (the most stable structure)
or as diamond, but the richness of its phase diagram does not end
there. In fact, many other structures have been proposed during the
past years, especially since experimental data suggested the existence
of a super hard phase of carbon.  Evidences for a structural phase
transition in compressed graphite to this unknown phase of carbon have
been reported in numerous
experiments~\cite{bundy_hexagonal_1967,goncharov_graphite_1989,hanfland_graphite_1989,utsumi_light-transparent_1991,zhao_x-ray_1989,yagi_high-pressure_1992,mao_bonding_2003}.
In fact, in the range of 10 to 25\,GPa one observes an increase of the
resistivity~\cite{bundy_hexagonal_1967} and of the optical
transmittance~\cite{goncharov_graphite_1989,hanfland_graphite_1989}, a
marked decrease of the optical
reflectivity~\cite{utsumi_light-transparent_1991}, changes in near
$k$-edge spectra~\cite{mao_bonding_2003} and in X-ray diffraction
(XRD)
patterns~\cite{zhao_x-ray_1989,yagi_high-pressure_1992,mao_bonding_2003}.
Several hypothetical structures have been proposed to explain these
features, such as hybrid $sp^2$--$sp^3$ diamond-graphite
structures~\cite{ribeiro_structural_2005},
M-carbon~\cite{li_superhard_2009},
bct-C$_4$-carbon~\cite{umemoto_body-centered_2010} and
W-carbon~\cite{wang_low-temperature_2011}. However, none of these
structures is able to match all experimental data in an unambiguous
and fully satisfactory manner.

\begin{figure}[t]
\setlength{\unitlength}{1cm}
\subfigure[]{\includegraphics[width=1\columnwidth,angle=0]{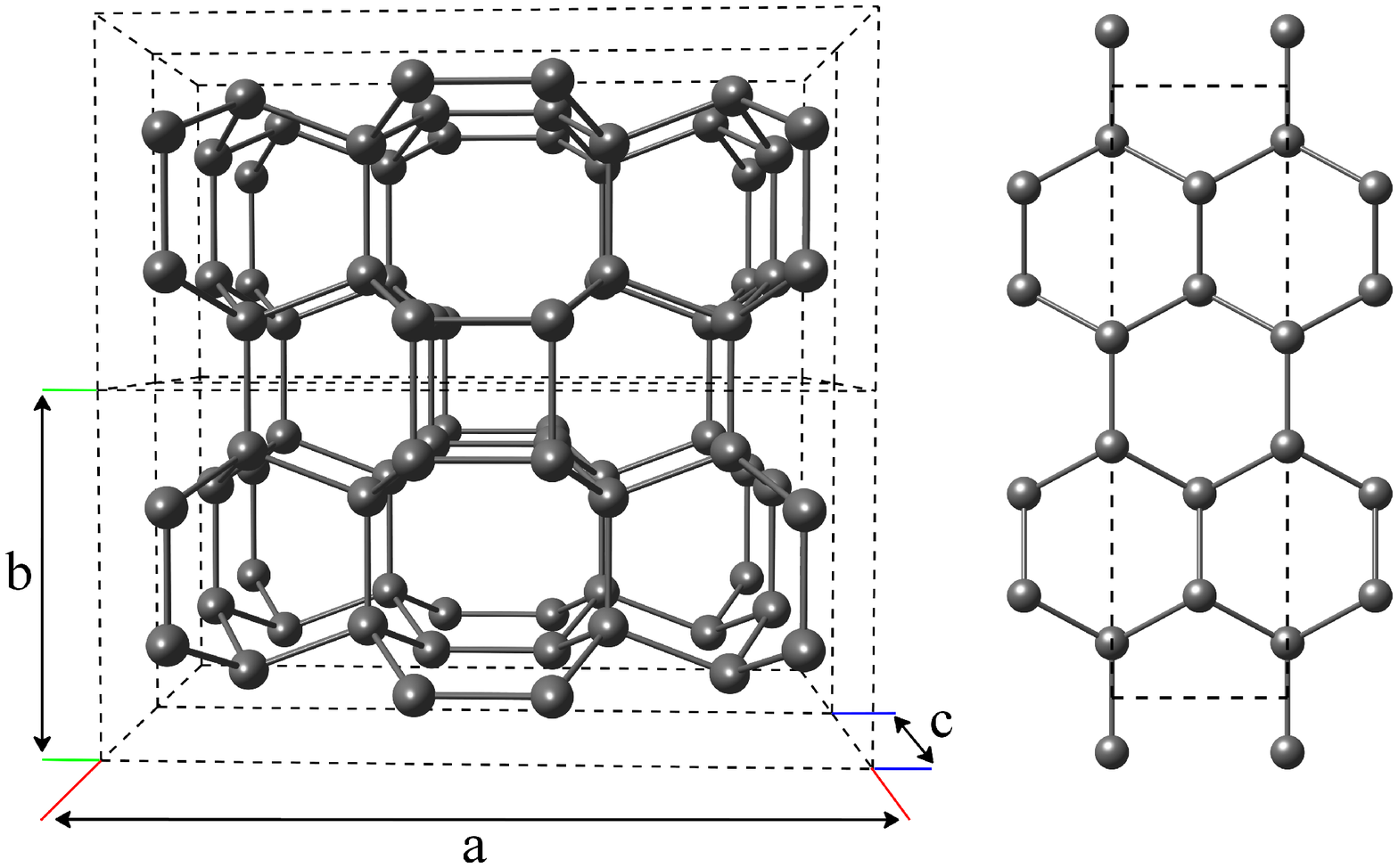}} 
\subfigure[]{\includegraphics[width=1\columnwidth]{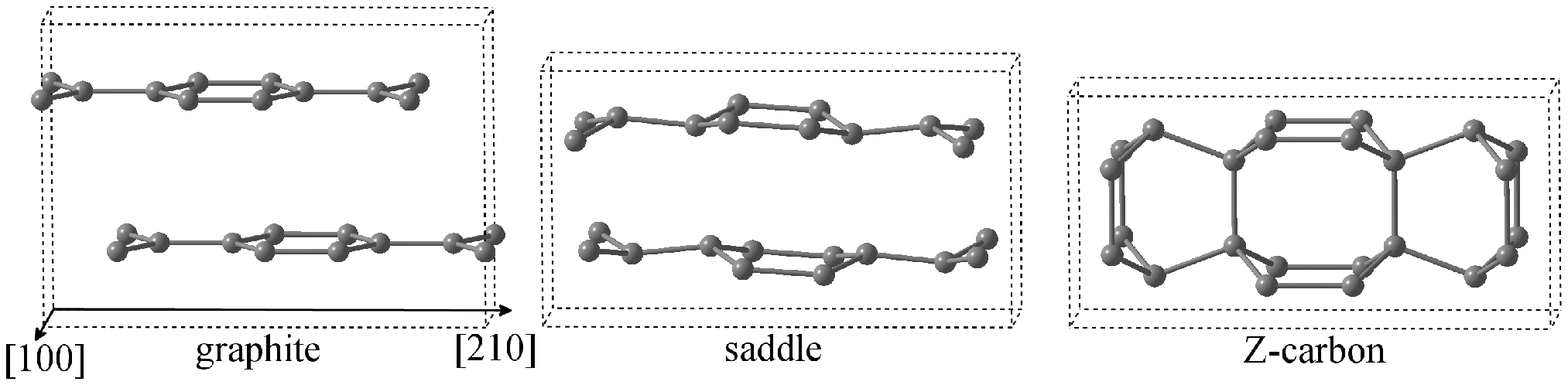}}
\caption{(a) Structure of Z-carbon viewed from two different angles revealing planar four-membered and 
non-planar eight-membered rings forming chains along the $b$-direction and channels in the $c$-direction. 
The graphene sheets are in the $a$-$c$ plane. (b) Proposed transition pathway from graphite to Z-carbon.}
\label{fig:structure}
\end{figure}

A common way to search for new crystal structures is to perform a
systematic survey of the enthalpy surface using some sophisticated
structure prediction method (for discussion on such methods see
Ref.~\cite{oganov_modern_2010}).  Here we use the minima hopping
method~\cite{goedecker_minima_2004}~(MHM) for crystal structure
prediction~\cite{amsler_crystal_2010}, which was designed to explore
low-enthalpy phases of materials. This method was coupled to the
all-electron projector-augmented wave method as implemented in the
{\sc abinit} code~\cite{gonze_brief_2005,bottin_large-scale_2008}.
Within the MHM, the system is moved from one configuration to the next
by performing consecutive molecular dynamics escape steps and geometry
relaxations.  The initial velocities for the dynamics are aligned
preferably along soft-mode directions in order to favor the escape to
low enthalpy structures.  Revisiting already known structures is
avoided by a feedback mechanism. Relaxations are performed by the fast
inertia relaxation engine~\cite{bitzek_structural_2006}.  The local
density approximation was employed based on its good description of
graphite. However, the enthalpy ordering was reconfirmed within the
generalized gradient approximation using two different functionals
(PBE~\cite{perdew_generalized_1996} and
PBEsol~\cite{perdew_restoring_2008}).  The most promising candidate
structures were then re-relaxed using norm conserving
Hartwigsen-Goedecker-Hutter
pseudopotentials~\cite{hartwigsen_relativistic_1998}.  Carefully
converged Mankhorst-Pack $k$-point meshes were used together with a
plane wave cut-off energy of 2100\,eV.

The MHM was employed using simulation cells with 4 and 8 carbon atoms
at a constant pressure of 15\,GPa. We found, in addition to previously
proposed structures of cold compressed graphite, a carbon phase that
we call Z-carbon.  This structure has {\em Cmmm} symmetry (see
Fig.~\ref{fig:structure}a) and, like diamond, is composed of $sp^3$
bonds. The conventional unit cell has 16 atoms with cell parameters at
0\,GPa of $a=8.668$\,\AA, $b=4.207$\,\AA, and $c=2.486$\,\AA, yielding
a cell volume of $V_0=90.7$\,\AA$^3$. The two inequivalent carbon
atoms occupy the $8p$ and $8q$ crystallographic sites with coordinates
$(1/3,y,0)$ and $(0.089,y,1/2)$, where $y=0.315$.  The structure
contains four-, six- and eight-membered rings, where planar
four-membered rings and non-planar eight-membered rings join together
buckled graphene sheets.  This structure can be interpreted as a
combination of hexagonal diamond and
bct-C$_4$-carbon~\cite{baughman_carbon_1997}.

In contrast to other structure prediction methods like evolutionary
algorithms or random search, the efficient escape moves in the MHM are
based on fundamental physical processes. Therefore, minima found
consecutively during a MHM simulation are usually connected through
low enthalpy barriers. Since we have observed escape moves to and from
Z-carbon to occur exclusively from and to graphite, we expect this
transition to be the most probable.  In Fig.~\ref{fig:structure}b we
show a possible transition pathway from graphite to Z-carbon. This
process is a combination of sliding and buckling of the graphene
sheets. The naturally staggered, i.e.  $AB$ stacked, graphene sheets
slide along the $[210]$ direction to an aligned $AA$ stacking while
the inter-layer distance decreases, and the aligned graphene sheets
deform to create an alternating armchair-zigzag buckling.

\begin{figure}[t]  
\setlength{\unitlength}{1cm}
\includegraphics[width=1\columnwidth,angle=0]{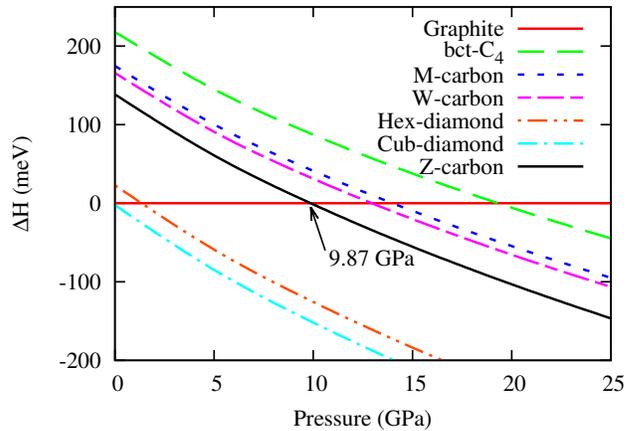}
\caption{Calculated enthalpy difference per atom with respect to graphite of
  several carbon allotropes as a function of pressure. Graphite is
  the horizontal line at zero. Z-carbon becomes more stable than
  graphite at around 10\,GPa. }
\label{fig:enthalpy}
\end{figure}

In order to investigate the relative stability of Z-carbon, the
calculated enthalpy difference with respect to graphite of several
allotropes are compared in Fig.~\ref{fig:enthalpy} as a function of
pressure. Z-carbon has the lowest enthalpy among all proposed
cold-compressed graphite phases, becoming more stable than graphite at
9.9\,GPa (around 2.5\,GPa below $W$-carbon).

We further investigated the dynamical lattice stability of this phase
by computing the phonon dispersion in the whole Brillouin zone. We
used linear-response theory in the framework of density functional
perturbation theory~\cite{gonze_dynamical_1997} with the {\sc abinit}
code. A proper convergence was ensured with a 12x12x12 $k$-point
sampling, a 4x4x4 $q$-point mesh, and a cut-off energy of 800\,eV.
All phonon modes were real confirming the strutural stability of the
structure.  Furthermore, from a fit of the Murnaghan equation we
obtained a bulk modulus of $B_0=441.5$\,GPa, and using the method
proposed by Gao~\etal~\cite{gao_hardness_2003} we calculated a
Vicker's hardness of $H_v=95.4$\,GPa.  Both bulk modulus and hardness
are extremely high and very close to the values for diamond
($B_0^{\textrm{diamond}}=463.0$\,GPa and
$H_v^{\textrm{diamond}}=97.8$\,GPa), which is compatible with the
observed ring cracks in diamond anvil cells~\cite{mao_bonding_2003}.

To investigate the energy gap of this material we used the
perturbative many-body GW technique starting from the local density
approximation~\cite{aulbur00}. These calculations reveal that Z-carbon
is an indirect band-gap material with a gap of around 4.7\,eV.
Therefore, this material is expected to be optically transparent in
agreement with
experiments~\cite{goncharov_graphite_1989,hanfland_graphite_1989}.


\begin{figure}[t]         

\setlength{\unitlength}{1cm}
\includegraphics[width=0.9\columnwidth,angle=0]{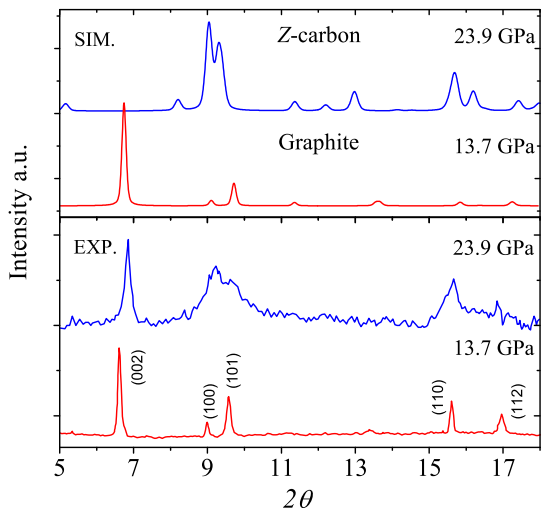}
\caption{Experimental XRD for cold compressed graphite at two
  different pressures from Ref.~\cite{mao_bonding_2003}
  and simulated XRD pattern for Z-Carbon (at 23.9\,GPa) and Graphite (at 13.7\,GPa). 
  The main characteristics of the proposed carbon are perfectly in
  agreement with the experimentally observed changes.}
\label{fig:xrd}
\end{figure}

We have gathered a considerable amount of experimental evidence that
points to the presence of Z-carbon in cold compressed graphite
samples. The first comes from the XRD experiment of
Ref.~\cite{mao_bonding_2003}.  In Fig.~\ref{fig:xrd} we can see that
the broadening of the XRD-spectra at high pressure can be explained by
the coexistence of graphite and Z-carbon.  However, the experimental
curve can also be explained to some extent by the other proposed
carbon
allotropes~\cite{li_superhard_2009,umemoto_body-centered_2010,wang_low-temperature_2011}
so that this experiment alone is not conclusive.

Other strong signatures for Z-carbon can be gathered from our
measurements of Raman spectroscopy under pressure. These experiments
were carried out at 300\,K using the 514.5\,nm line excitation of an
Ar$^+$ laser, and a Jobin-Yvon HR-800 Labram spectrometer with
double-notch filtering with resolution better than 2\,cm$^{-1}$.  In
the high pressure Raman measurements, we used a diamond anvil cell to
apply pressure on two different samples (single crystals of graphite
and highly oriented pyrolitic graphite), inside a 120 micron hole
drilled in an iconel gasket. Argon and paraffin was used as the
pressure medium.  The pressure was determined by the ruby luminescence
of a small chip ($<30$ microns).  The laser was focused down to 3
microns with a power of about 20\,mW on the sample.

\begin{figure}[t]      
\setlength{\unitlength}{1cm}
\includegraphics[width=0.9\columnwidth]{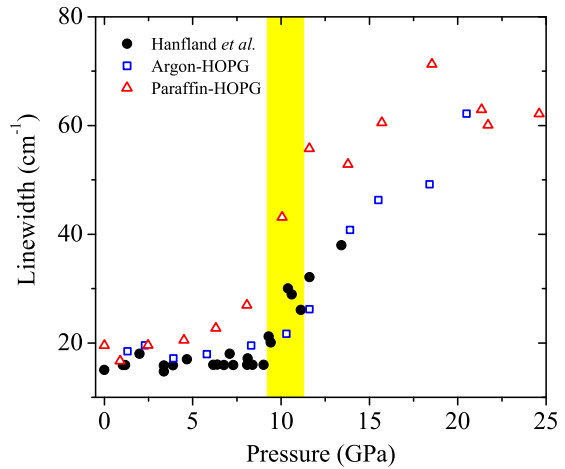}
\caption{Experimental linewidth of the G-band of graphite 
under pressure. The linewidth stays nearly constant until 
pressures of the order of $9-10$\,GPa, above which the linewidth 
begins to broaden rapidly. This is a strong evidence for a 
structural transition in graphite. Experiments were conducted using  
highly oriented pyrophillitic  
graphite (HOPG) and argon (squares) or paraffin oil (triangles) as 
pressure transmitting media. 
The black dots are taken from Ref.~\cite{hanfland_graphite_1989}}.
\label{fig:gband}
\end{figure}

The principal Raman active mode of graphite is the G-band at 1579
cm$^{-1}$ (at 0\,GPa) which originates from the $sp^2$ carbon atoms
vibrating in-plane with $E_{2g}$ symmetry. The effect of hydrostatic
pressure on the linewidth of the G-band is shown in
Fig.~\ref{fig:gband}.  The linewidth remains nearly constant until
around 9--10\,GPa.  Above this value, the linewidth begins to broaden
rapidly, in agreement with previous results of Hanfland
\etal~\cite{hanfland_graphite_1989}.  (A similar broadening has also
been reported for turbostratic graphite-like BC$_4$ under
pressure~\cite{solozhenko_raman_2007}.) This behavior is a sign of a
structural transformation at this pressure, and can be explained by
important changes in the Raman cross section caused by interlayer
coupling and the formation of $sp^3$ bonds. As seen in
Fig.~\ref{fig:enthalpy}, Z-carbon becomes enthalpically favored with
respect to graphite at around 10\,GPa, whereas all other proposed
structures cross the graphite line at significantly higher pressures.

\begin{figure}[t] 
\setlength{\unitlength}{1cm}
\includegraphics[width=0.9\columnwidth,angle=0]{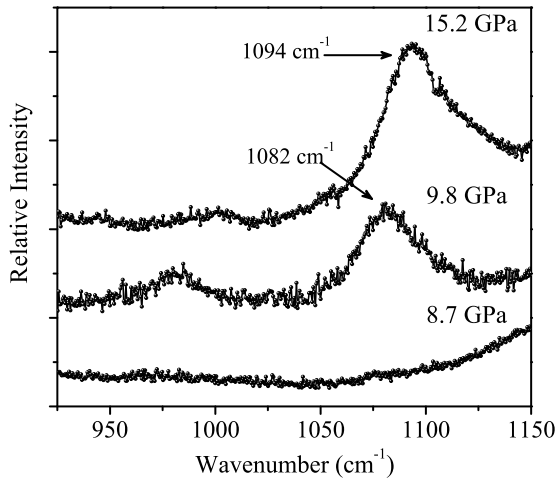}
\caption{Experimental Raman spectra of graphite under pressure. 
  The peak around 1082 cm$^{-1}$ appearing at around 10\,GPa and
  its evolution can be explained by either Z-carbon or W-carbon.}
\label{fig:raman}
\end{figure}

Finally, we present the strongest direct evidence of the existence of
Z-carbon, which is found in the Raman spectrum of graphite under
hydrostatic pressure, shown in Fig.~\ref{fig:raman} for the energy range below the $1^{st}$ order Raman peak of 
diamond (1332 cm$^{−1}$ at 0GPa)~\cite{occelli_properties_2003}.  Neither graphite nor
cubic-diamond have Raman active peaks in the selected energy region,
however we can observe that a clear peak appears at 1082\,cm$^{-1}$
for pressures higher than 9.8\,GPa. This peak can not be explained by
either bct-C4 carbon, M-carbon, nor by the pressure medium (argon). 
The only structures that have Raman active modes
compatible with this experimental evidence are Z-carbon and W-carbon.
For Z-carbon the frequencies are 1096.5\,cm$^{-1}$ at 10\,GPa and
1110\,cm$^{-1}$ at 15\,GPa.  Incidentally, Z-carbon also has a Raman
active $A_g$ mode at 1348.5\,cm$^{-1}$ at 0\,GPa (theoretical value).
This appears as a signature of planar four-membered rings that
overlaps with the so-called defect D-band of graphite at around
1345.5\,cm$^{-1}$ at 0\,GPa (experimental value).

In conclusion, we identified an allotropic structure of carbon,
Z-carbon, that becomes more stable than graphite above 10\,GPa.  From
all known carbon allotropes, only cubic and hexagonal diamond have
lower enthalpy at high pressures. The Z-carbon structure is as
hard as diamond, and is transparent in the optical region. Moreover, a
wide range of experimental data can only be explained by the presence
of Z-carbon in samples of cold compressed graphite: first, the
features of the X-ray diffraction spectra of graphite under pressure
exhibit a broadening that matches the main peaks of Z-carbon. Second,
the principal Raman signal of graphite, the G-band mode, suffers an
abrupt increase of the linewidth above 9-10\,GPa --- the pressure
range where Z-carbon becomes more stable than graphite. Third, a new
peak at 1082 cm$^{-1}$ appears in the Raman spectrum of graphite at
around 10\,GPa, at the frequency of a Raman active mode of Z-carbon.
The only candidate among all carbon allotropes proposed so far that
can explain all above features simultaneously is Z-carbon. 

Our work also highlights the promising prospects of the minima hopping
method for crystal structure prediction~\cite{amsler_crystal_2010}.
The exploration of the structural variety of even simple elements such
as carbon was up to now typically the subject of many different
studies which were presented in numerous papers over many years.  In
this first application of the MHM we were able to find not only
Z-carbon, but also all other known carbon phases at the given pressure
condition fully automatically. We can therefore expect 
that this method can also find with high reliability the low energy
structures of many other materials for which our knowledge is at
present still rudimentary, leading to important advances 
in the field of solid state physics.

\acknowledgments We thank T.\,J. Lenosky and A.\,R. Oganov for
valuable discussions. We thank Gilles Montagnac and H\'erv\'e Cardon,
from the Laboratoire de Geologie de Lyon, France, for technical
support during the Raman experiments. Financial support provided by
the Swiss National Science Foundation is gratefully acknowledged.
JAFL acknowledges the CONACyT-Mexico. SB acknowledges support from
EU’s 7th Framework Programme (e-I3 contract ETSF) and MALM from the
French ANR (ANR-08-CEXC8-008-01). Computational resources were
provided by the Swiss National Supercomputing Center (CSCS) in Manno
and IDRIS-GENCI (project x2011096017) in France.

MA and JAFL contributed equally to this work.

%

\end{document}